\documentclass[aps, prl, preprint]{revtex4-1}

\usepackage{amsmath,amssymb, amsfonts}
\usepackage{graphicx}

\def\braket#1{\mathinner{\langle{#1}\rangle}}
\newcommand{\sbraket}[1]{\lbrack #1\rbrack}

\newcommand{\ii}{\mathrm{i}}

\newcommand{\al}{\alpha'}

\newcommand{\aord}[1]{\mathcal{O}\left(\alpha'^{#1} \right)}

\begin{document}

\title{Maximal R-symmetry violating amplitudes in type IIB superstring theory} 
 
\author{Rutger H.  Boels}
\email{Rutger.Boels@desy.de}
\affiliation{II. Institut f\"ur Theoretische Physik,\\ Universit\"at Hamburg,\\  Luruper Chaussee 149,\\ D-22761 Hamburg, Germany}

\date{\today}

\begin{abstract}
\noindent On-shell superspace techniques are used to quantify R-symmetry violation in type IIB superstring theory amplitudes in a flat background in ten dimensions. This shows the existence of a particularly simple class of non-vanishing amplitudes in this theory which violate R-symmetry maximally. General properties of the class and some of its extensions are established which at string tree level are shown to determine the first three non-trivial effective field theory contributions to all multiplicity. This leads to a natural conjecture for the exact analytic part of the first two of these.
\end{abstract}





\maketitle

\section{Introduction}
Scattering amplitudes are central objects in the study of physics at high energies as they form a measure for the probability of a certain scattering event taking place. They are therefore a stepping stone in the bridge from theory to experiment. Textbook methods exist in principle to calculate them in string and field theory. Explicit calculations of amplitudes of especially closed strings in ten dimensions are however limited to partial results for five and six particles in a flat background due to the complexity of the computations and results. For open strings only very recently some progress has been made \cite{Mafra:2011nw}. This holds despite the fact that its amplitudes are counted as one of the best understood aspects of string theory. 

It invariably pays to make as much symmetries of the theory manifest in the amplitudes as possible since this tends to lead to the most compact expressions. Vice-versa, if compact expressions are found after a long calculation an un-accounted for symmetry is naturally suspected. A prime example of this are Maximally Helicity Violating (MHV) scattering amplitudes in four dimensional gauge and gravity theories which after summing over large numbers of Feynman graphs gives a single line expression \cite{Parke:1986gb} at tree level. With all particles incoming, MHV amplitudes are those for which the difference between number of positive ($n_+$) and negative ($n_-$) helicity particles is maximal: it saturates the bound $|n_+ - n_-|\geq n_+ + n_- - 4 $. Tree amplitudes beyond this bound vanish, even  to all loop orders in supersymmetric theories \cite{Grisaru:1977px}. The MHV amplitude and its hidden symmetries are central to many recent exciting developments in four dimensions, see \cite{Dixon:2011xs} and its references. 

MHV amplitudes are special however to four dimensions. No similarly simple series of amplitudes is known above four dimensions. This is a bit of a puzzle for superstring theories which naturally live in ten dimensions: since these are much more symmetric than gauge theories one would expect their amplitudes to be even simpler. Clearly, a new and efficient ordering principle is needed.

We propose in this article that an on-shell superspace presented in \cite{Boels:2012ie} can serve as such a principle for the type IIB closed superstring in a flat background. This superspace can be used to solve the on-shell supersymmetric Ward identities systematically. The simplest solution to the identities will lead to a ten-dimensional analog of the four dimensional MHV amplitudes. Instead of violating helicity maximally, these violate R-symmetry maximally.

\section{On-shell superspace for amplitudes}
An on-shell massless momentum can be written \cite{Boels:2012ie}  in terms of Weyl spinors solving the massless Dirac equation as
\begin{equation}\label{eq:spinornorm}
k_{\mu} \sigma^{\mu,BA'} = \lambda^{B, a}  \lambda^{A'}_{a}  \, ,
\end{equation} 
The capital Roman indices indicate are $16$ dimensional Weyl spinor indices, while the lower-case roman indices indicates the $8$ dimensional little group Weyl spinors. The summation convention is understood. The supersymmetry algebra is
\begin{equation}
\{ \overline{Q}^B,Q^{A'} \} = k^{BA'}  \, ,
\end{equation} 
which on-shell by the equations above can be written as
\begin{equation}\label{eq:susyonshell}
\{ \overline{Q}^B,Q^{A'} \} =  \lambda^{B, a}  \lambda^{A'}_{a}  \, .
\end{equation} 
It is easy to check that for fermionic $\eta$ variables
\begin{equation}\label{eq:defsuperspacegens}
\overline{Q}^B = \lambda^{B, a}   \frac{\partial}{\partial \eta^a}    \qquad Q^{A'}= \lambda^{A'}_{a} \eta^a \, ,
\end{equation}
is a representation of the massless on-shell supersymmetry algebra in equation \eqref{eq:susyonshell} since
\begin{equation}
\{\eta^{a},    \frac{\partial}{\partial \eta^b}    \} = \delta^a{}_{b}  \, ,
\end{equation}
holds for fermionic variables. The chiral on-shell supersymmetry algebra has a $U(1)_R$ symmetry rotating $Q$ and $\overline{Q}$. The variable $\eta$ has a natural charge which will be taken to be $1$. Every leg of a scattering amplitude can now be given as a function of the on-shell superspace variables $(k,\eta)$, 
\begin{equation}
\phi(k,\eta) = \phi^0(k) + \phi_{a}(k) \eta^{a} + \frac{ \phi_{ab}(k)}{2!}\eta^{a} \eta^b \ldots + \left(\eta\right)^{8} \overline{\phi^0}(k) \, .
\end{equation}
As explained elsewhere  \cite{Boels:2012ie}  \cite{Boels:2012if}, the little group transformations of the field $\phi^0$ determine the field content. Here this field is the holomorphic scalar $\tau = a + \ii e^{-\phi}$ which makes this the massless multiplet of type IIB supergravity  \cite{Boels:2012if}. The field $\phi^0$ has $U(1)_R$ charge four in our conventions. The component fields then have natural charges $4-f$ where $f$ is the number of fermionic variables. Gravitons are located at fermionic weight four.

\subsection{On-shell susy Ward identities and their solution}
The amplitude with $n$ particles can now be promoted to a function on $n$ copies of the on-shell superspace. The on-shell supersymmetry Ward identities have a compact expression on these  superamplitudes,
\begin{equation}
 \overline{Q}^B A_n \,=\, 0\, =\, Q^{A'} A_n
\end{equation}
where $Q$ and $\overline{Q}$ are the sum over the supersymmetry generators for each leg separately. These equations are exact: they only depend on the on-shell supersymmetry algebra, not on any possible coupling constant. In the representation of equation \eqref{eq:defsuperspacegens} $Q^{A'}$ is a pure multiplication operator. The only non-trivial way for this operator to annihilate an amplitude is if the amplitude is proportional to the fermionic delta function,
\begin{equation}
\delta^{16}(Q) \equiv \frac{1}{16!} \epsilon_{A'_1 \ldots A'_{16}} Q^{A'_1} \ldots Q^{A'_{16}} \, .
\end{equation}
Since this function is a polynomial of Q, the action of  $\overline{Q}$ simply results in an expression proportional to the overall momentum. Hence a general super amplitude can be written as 
\begin{equation}\label{eq:genampexp}
A_n = \delta(K) \delta(Q) \tilde{A}_n \quad \textrm{with} \quad \overline{Q}^{A} \tilde{A} = 0  \, .
\end{equation} 
with $\delta(K)$ the momentum conserving delta function. There are no solutions with less fermionic weight than $16$, apart from the kinematically special case of three particles for which a weight $12$ solution exists. There is also a maximal weight: this is obtained from the minimal weight on the conjugate superspace where the fermionic multiplication and differentiation in  equation \eqref{eq:defsuperspacegens} are interchanged. In general
\begin{equation}\label{eq:minmaxweight}
16 \leq \textrm{weight} \leq 8 n  -  16
\end{equation}
holds with an even fermionic weight. For the kinematical exception  of three massless legs a solution to the Ward identities exists which has fermionic weight $12$,
\begin{equation}\label{eq:threepointexcep}
A_{3}(G,G,G) \propto \delta^{12}(Q) 
\end{equation}
see \cite{Boels:2012ie} for the explicit expression. Note component amplitudes with only gravitons on external legs are located at the ``middle'' fermionic weight $4 n$. 

\subsection{Maximal R-symmetry violation}
The extremal fermionic weight of superamplitudes leads directly to vanishing component amplitudes: those obtained by expanding a generic superamplitude with less than $16$ or more than $8 n -16$ fermionic variables. For example 
\begin{equation}
A_n(\phi,\phi,\phi, \ldots ,\phi) =0 \qquad A_n(\bar{\phi},\phi,\phi, \ldots ,\phi) =0 
\end{equation}
hold as well as their complex conjugates. Note the vanishing of these amplitudes is exact. The simplest class of non-vanishing superamplitudes are those which have fermionic weight $16$ and hence must be proportional to the delta function. This class encompasses all component amplitudes which are obtained by $16$ fermionic integrations. For instance, 
\begin{equation}\nonumber
A_n(\bar{\phi},\bar{\phi},\phi, \ldots ,\phi)  = \tilde{A} \int d\eta_1^{8}  d\eta_2^{8} \delta^{16}(Q)  = \tilde{A} \left(2 k_1 \cdot k_2 \right)^{4}
\end{equation}
This amplitude violates the $U(1)_R$ symmetry maximally, by $4 n - 16$ units. In general a superamplitude written as
\begin{equation}
A_n = (\kappa)^{n-2} \delta(Q) \tilde{A}_n 
\end{equation}  
which includes an overall tree level power of the gravitational coupling constant  $\kappa$ with weight
\begin{equation}
[\tilde{A}_n] = 2 l  \qquad l \in \{0,1, 4 (n - 4)\} \, .
\end{equation}
will violate $U(1)_R$ by $4 n - 16 + 2 l$ units.  The amplitudes with minimal fermionic weight $16$ which are proportional to the fermionic delta function will be called ``Maximally R-symmetry Violating'' (MRV). A conjugate amplitude with maximal fermionic weight also exists: this has minimal weight on the conjugate superspace. The four and three particle amplitudes with massless legs are exceptional as they always preserve the symmetry; their concrete expressions are in  \cite{Boels:2012ie}.

The analysis above can be extended to massive matter, taking into account the massive on-shell superspace in ten dimensions constructed in  \cite{Boels:2012if}.

\section{Properties of MRV amplitudes}
For MRV amplitudes the function $\tilde{A}$ in equation \eqref{eq:genampexp} contains no fermionic variables and is a completely symmetric function of the external momenta by Bose symmetry. Moreover, it cannot have massless poles for more than four particles: on such a pole the amplitude would factorize by tree level unitarity,
\begin{equation}
\lim_{k_{ij}^2 \rightarrow 0} \tilde{A}_n \delta^{16}(Q) \stackrel{?}{\rightarrow} \int d\eta^8 A_L \,\frac{1}{k_{ij}^2} \, A_R 
\end{equation}
where $k_{ij}$ stands for any kinematic invariant. The minimal fermionic weight of scattering amplitudes is $16$, with the exception of the three point amplitude which has weight $12$. The only way the $8$ fold fermionic integral gives a fermionic weight $16$ answer is for the four point amplitude. By the same weight counting argument the fermionic weight $18$ amplitudes also do not have massless poles, while fermionic weight $20$ and $22$ amplitudes can only have poles where two legs become collinear, the latter only when $n>5$. It can be shown MRV amplitudes can have supersymmetric residues at massive poles. 

A non-trivial kinematic limit for closed string amplitudes is the ``soft'' limit where one momentum and it's superpartner go to zero, $k_i,\eta_i \rightarrow 0$. This reduces to a limit of an axion and a dilaton leg on a superamplitude. The axion part vanishes in the $k_i \rightarrow 0$ limit because there are only field strength couplings in the vertex operators. For the soft limit on the dilaton one obtains \cite{Ademollo:1975pf}\cite{Shapiro:1975cz} for a closed string amplitude with massless legs
\begin{equation}\label{eq:softlimit}
 \lim_{k_1\rightarrow 0} A_{n+1}(\{k_1\}, X) =    g_s \alpha'^{2} \left(  \alpha' \frac{\delta}{\delta \alpha'} -  2 g_s   \frac{\delta}{\delta g_s}  \right) A_{n}(X)
\end{equation} 
with $g_s$ the closed string coupling constant. This follows from rewriting the dilaton vertex operator in the soft limit as a variation of the string action. By construction the differential operator annihilates the gravitational coupling constant $\kappa \equiv g_s (\alpha')^{2}$. 

A final constraint is the behavior under a so-called supersymmetric BCFW shift  \cite{Britto:2005fq} \cite{Boels:2012ie}. The bosonic part of this shift changes the momenta of two selected legs of the superamplitude as
\begin{equation}
k_i \rightarrow k_i + q z \qquad k_j \rightarrow k_j - q z
\end{equation}
for a $q$ such that $q^2 = q \cdot k_i = q \cdot k_j = 0$. From   \cite{Cheung:2010vn}, \cite{Boels:2010bv}  it follows that under this shift on two massless superfield legs the string theory superamplitude should shift as
\begin{equation}
\lim_{z\rightarrow \infty} A \propto (z)^{-2 \al k_i \cdot k_j - 2} 
\end{equation}
For MRV amplitudes this is the behavior under a bosonic shift of the bosonic pre-factor $\tilde{A}$. 

\section{Effective field theory expansion}
In the effective field theory expansion of string theory all dimensionless kinematic invariants such as $\alpha' k_i \cdot k_j$ are taken to be small and hence the only propagating degrees of freedom are massless. In this regime the MRV amplitudes must be local expressions expressible as a power series in $\alpha'$. From dimensional analysis it follows that
\begin{equation}\label{eq:lowenergyexpansion}
\tilde{A}_n^{\textrm{MRV}} =  \alpha'^3 c_0 +  \alpha'^4 c_1 +  \alpha'^5 c_2  + \aord{6} 
\end{equation}
with the coefficients $c_i$ completely symmetric polynomia of the external kinematic variables of mass dimension $2i$. Hence $c_0$ is a constant and $c_1$ vanishes as it must be proportional to 
\begin{equation}
c_1 \sim (k_1 + k_2 + \ldots k_n)^2 = 0
\end{equation} 
The coefficient $c_2$ can be a linear combination of three polynomials in different orbits of the symmetrization. There are two possible completely symmetric constraints derived from momentum conservation at this order, leaving only one independent polynomial above three points. Similarly, at order $\alpha'^6$ there are $2$ independent polynomials above four points.

As a concrete example consider the MRV amplitude at five points. This closed string superamplitude can be calculated from the component amplitude $A(g^{++}, g^{++}, g^{--}, g^{--}, \phi)$ in four dimensional kinematics. The latter follow from known results about open string amplitudes \cite{Stieberger:2006te} and the KLT relations \cite{Kawai:1985xq}, see also \cite{Elvang:2010kc}. This component amplitude also follows by fermionic integration over the fermionic delta function which in this case yields an easily identifiable  $\braket{12}^4 \sbraket{34}^4$ spinor factor which is  $R^4$ on-shell. Expanding the resulting complicated expression using Mathematica with the package HypExp \cite{Huber:2007dx} yields:
\begin{multline}\nonumber
\tilde{A}_5^{\textrm{MRV}} = -6\, \zeta(3) \alpha'^3 - \frac{5}{2^5}\, \zeta(5)  \alpha'^5 \left([s_{12}^2]_5  \right) +\\ \frac{1}{2^5} \,\zeta(3)^2 \alpha'^6 \left([s_{12}^3]_5\right) - \frac{7}{2^{13}} \zeta(7) \, \alpha'^7 \left(13[s_{12}^4]_5 + 6 [s_{12}^2 s_{34}^2]_5 \right) +  \\ \frac{1}{15} \frac{1}{2^{11}} \zeta(3)\zeta(5) \,\alpha'^8 \left(71 [s_{12}^5]_5 + 25[s_{12}^3 s_{34}^2]_5 \right) +  \aord{9}
\end{multline}
for the MRV amplitude where the brackets indicate sums over all permutations of the legs normalized such that the resulting polynomial has terms with coefficient one. For instance, 
\begin{equation}
[s_{12}^j]_n = \frac{1}{2}\,\frac{1}{(n-2)!} \sum_{\sigma \in \textrm{perms}(1,2,\ldots, n)} s_{\sigma_1 \sigma_2}^j
\end{equation}
where $s_{ij} = (k_i + k_j)^2$. Note the appearance of odd $\zeta$ values only which have at order $\alpha'^i$ transcendental weight $i$ as first observed for certain graviton amplitudes in \cite{Stieberger:2009rr}.

\subsection{Tree amplitudes to all multiplicity}
There is a neat interplay between the low energy expansion of equation \eqref{eq:lowenergyexpansion} and the soft limit \eqref{eq:softlimit}. The completely symmetric polynomials of the external momenta simply reduce from the $n$-particles to the $n-1$ particle case. Hence there are interrelations between superamplitudes with different numbers of external particles. These relations can be degenerate. At five and higher points for instance there are two independent symmetric polynomial of the external momenta at mass dimension $6$, while there is only one at four points. 

For the first three coefficients in equation \eqref{eq:lowenergyexpansion} degeneracy is not a problem: for these an all-multiplicity result can be derived from the known 4 and 5 point amplitudes,
\begin{equation}\label{eq:MRVexptreelevel}
\tilde{A}_n^{\textrm{MRV}} =  2 \cdot 3^{n-4} \alpha'^3 \zeta(3)+  \frac{5^{n-4}}{2^5} \alpha'^5 \zeta(5) \left([s_{12}^2]_n  \right) +   \frac{6^{n-4}}{3 \cdot 2^6} \, \alpha'^6 \zeta(3)^2 \left([s_{12}^3]_n\right) +  \aord{7} 
\end{equation}
at string tree level.

\subsection{Exacting amplitudes}
There is a large body of work on the structure of the string theory effective action in type IIB and the symmetries which are preserved. Since MRV amplitudes in the effective field theory expansion are local, there should be a close connection. Most of this will be left to future work save a natural conjecture.

Type IIB string theory has an exact $\textrm{SL}(2, \mathbb{Z})$ symmetry. This manifests itself for instance in the effective action of type IIB in the on-shell pre-factors of local terms such as $R^4$ which have this symmetry and are consistent with all known data. Extremal examples of this series are the $R^4$ term which controls four graviton scattering and the $\lambda^{16}$ term which controls  the scattering of $16$ holomorphic dilatinos. The latter term violates R-symmetry by $48$ units, while the former preserves it. Hence these terms are closely related to MRV amplitudes.

The coupling constant dependence of these terms can be given in Einstein frame as a certain derivative of a non-holomorphic Eisenstein series \cite{Green:1997tv} \cite{Green:1997me}. The result can be written in terms of a modular covariant function of vacuum expectation values of the scalar fields,
\begin{equation}
f_{\beta}^{k}(\tau_b,\bar{\tau}_b) = \sum_{(l,m) \neq (0,0)} (l+ m \tau_b)^{k - \beta} (l+m \bar{\tau}_b)^{-k - \beta}
\end{equation}
where $\tau_b = \langle a\rangle + \ii e^{- \langle \phi \rangle } \equiv  \langle a\rangle  + \frac{\ii}{g_c}$. The weak string coupling expansion contains a tree level and a few  loop contributions plus an infinite series of D-instanton corrections. A term which violates $U(1)_R$ by $4i$ units has $k=i$ \cite{Green:1997me}. For the $R^4$ term for instance $\beta= \frac{3}{2}$ and $k=0$, while for the $\lambda^{16}$ term $\beta= \frac{3}{2}$ and $k=12$.   Generalizations to $\alpha'^5$  \cite{Green:1999pu} and $\alpha'^6$ \cite{Green:2005ba} are known.
 
From this known structure a simple conjecture follows for the exact form of the analytic part  (see \cite{Green:1997tv}) of MRV amplitudes for the first two orders in the effective field theory expansion
\begin{equation}\nonumber
\tilde{A}_n^{\textrm{MRV}} = 3^{n-4} \alpha'^3 f_{\frac{3}{2}}^{n-4} + \frac{5^{n-4}}{2^6} \alpha'^5   f_{\frac{5}{2}}^{n-4} \left([s_{12}^2]_n  \right) + \aord{6}
\end{equation}
for all multiplicity which reproduces the tree level results of equation \eqref{eq:MRVexptreelevel} in the limit $g_c \rightarrow 0$. A natural but more involved guess for the $\alpha'^6$ term can be formulated. 

\section{Conclusion and discussion}
In this Letter a new ordering of scattering amplitudes of the IIB superstring in a flat background according to their R-symmetry violating properties has been introduced. The amplitudes which violate R-symmetry maximally are the simplest, in close analogy with MHV amplitudes in four dimensions. In the effective  field theory expansion the symmetry constraints on these amplitudes determine them to a large degree, displaying remarkable simplicity. The obtained amplitudes are the first non-trivial all-multiplicity expressions in higher dimensional theories.  

Since amplitudes are such a central concept in string theory there are many avenues to pursue from here. Concrete results on beyond-MRV amplitudes are interesting as well an MRV analog of \cite{Stieberger:2012rq} for the integral basis of the tree amplitudes. The relation of the MRV amplitudes to the effective action as well as worldsheet methods need clarification as these are portals to string theory in curved backgrounds.

More generally MRV amplitudes have the potential to be as important to type IIB superstring theory as the concept of MHV amplitudes is to four dimensional gauge and gravity theories. This should shed new light on the structure of type IIB superstring theory and beyond. 

\section*{Acknowledgements}
This work has been supported by the German Science Foundation (DFG) within the Collaborative Research Center 676 ``Particles, Strings and the Early Universe" and in part by Perimeter Institute for Theoretical Physics.

\bibliographystyle{apsrev4-1}

\bibliography{MRV.bib}

\end{document}